\documentclass[aps,twocolumn,epsfig,graphics,showpacs,floatfix,mathbbm]{revtex4}

\usepackage{amsmath,amsfonts,amssymb,graphics,graphicx,epsfig,color,times,bbm,booktabs}

\begin{document}

\title{Cluster state preparation
using gates operating at arbitrary success probabilities}

\author{K.\ Kieling, D.\ Gross, and J.\ Eisert}

\address{Blackett Laboratory, Imperial College London, Prince Consort Road, London SW7 2BW, UK }
\address{Institute for Mathematical Sciences, Imperial College London, Prince's Gate, London SW7 2PG, UK}
\date\today

%\maketitle

\begin{abstract} 
Several physical architectures allow for
measurement-based quantum computing using
sequential preparation of
cluster states by means of
probabilistic quantum gates.  In such an approach, the order in which
partial resources are combined to form the final cluster state turns
out to be crucially important.  We determine the influence of this
classical decision process on the expected size of the final cluster.
Extending earlier work, we consider different quantum gates operating
at various probabilites of success.  For finite resources, we employ a
computer algebra system to obtain the provably optimal classical
control strategy and derive symbolic results for the expected final
size of the cluster.  We identify two regimes: When the success
probability of the elementary gates is high, the
influence of the classical control strategy is found to be negligible.
In that case, other figures of merit become more relevant. In 
contrast, for small probabilities of success, the choice of 
an appropriate strategy is crucial.
\end{abstract}


\maketitle

\section{Introduction}

Measurement-based quantum computing has a number of appealing features not present in the 
standard gate model. From an experimental perspective, it may well be a significant 
advantage to abandon the need for exact unitary control between any two constituents, 
and separate the process of entanglement generation from that of entanglement 
consumption. In such measurement-based computing, an entangled state is generated,
followed by a sequence of local measurements on single constituents. In the
original one-way computer  \cite{BR01}, this universal 
resource is the cluster state \cite{BR01,Survey}. 
In the following, we will indeed concentrate on cluster state preparation; 
see, however, Ref.\ \cite{New} for a method of constructing a number
of novel models for measurement-based computing that 
make use of resource states different from the cluster state.

Generally speaking, there are two ways of preparing cluster states: On the one hand, this can be
done by means of translationally invariant local interactions, not requiring individual
local control. This most 
prominently applies to preparations using 
cold collisions of atoms in optical lattices by applying spin-dependent shifts \cite{Cold,Greiner}.
The other way is to build up cluster states from {\it 
elementary building blocks},
such as entangled pairs, a framework we will concentrate on in this work. This is the setting 
that plays the key role when applied to a number of physical
architectures: Specifically, it is the preferable or 
typically only applicable method in preparations using linear 
optical systems \cite{KLM01,BR04,Zeilinger,BCZ+06,GKE06}, 
optical systems 
with weak non-linearities \cite{LNMS06},
trapped atoms \cite{Atoms}, and matter qubits in optical cavities 
\cite{Monroe,Kimble,Cavity3,Cavity4}.
Here, the quantum 
gates that are applied are typically inherently probabilistic, 
necessarily leading to a significant
overhead in required resources. 
For linear optical settings in particular,
the probabilistic character of quantum gates is 
unavoidable~\cite{Scheel,Eisert05}. 

There is a new element in this idea 
that was not present before: {\it Choice}. Indeed,
when building up resource states from smaller blocks, several kinds of
intermediate structures will appear, and it turns out to be crucial to make a meaningful choice 
of which parts to attempt to link at what stage. This is no 
marginal effect, but can give rise to 
differences in orders of magnitude in, say, 
consumption of maximally entangled pairs of qubits 
with state vectors of the form $|0,0\rangle+|1,1\rangle$ (
referred to as EPR pairs).
This would apply to maximally entangled photon
pairs in the optical context.  Such an advantage may be
gained even for moderately-sized cluster states~\cite{KGE06}.
Fortunately, for building up cluster chains using 
gates with an elementary success probability of 
$p_{\textrm s}>0$,
an overhead is sufficient that is linear in the size of 
the final chain \cite{KGE06,GKE06}.
For the special case of $p_{\textrm s}=1/2$ 
(as, e.g., in linear optical architectures) 
the consumption of five EPR pairs per 
edge constitutes an upper bound for the optimal strategy,
a bound that can not be improved any more with sequentially acting
gates operating at the success probability dictated by linear optics.
Further, for any $p_{\textrm s}>0$ the overhead required to produce a 2D cluster out of chains is also only
linear in the size of the state to produce. As all these processes are probabilistic, the quoted results actually state
that a constant overhead per site is sufficient for any $p_{\textrm s}>0$ to produce 
a state of size $n\times n$ (or $n$ in the 1D case) almost certainly as $n$ becomes 
large \cite{GKE06}.

\begin{table}
%  \begin{indented}\item[]\begin{tabular}{lccl} \br
  \begin{tabular}{lccp{6cm}} \hline
    Gate & $d_{\textrm s}$ & $d_{\textrm f}$ & Physical realization\\ \hline %\mr
    CZ      & 1 & 2 & Distant atoms~\cite{Atoms,Kimble} \\
    KLM CZ  & 1 & 1 & Linear optics with $p_{\textrm s}=1/16$~\cite{KLM01} or $p_{\textrm s}=1/4$~\cite{PJF01}; weak non-linearities with $p_{\textrm s}=3/4$~\cite{LNMS06};
                      linear optics with $p_{\textrm s}=p_{\textrm{NDM}}/9$~\cite{RLBW02,HT02}; $p_{\textrm s}=1/8$~\cite{BCZ+06} \\
    DPC     & 0 & 2 & Trapped atoms and frequency qubits~\cite{Monroe}, $p_{\textrm s}<1/4$ \\
    Fusion  & 0 & 1 & Linear optics parity check~\cite{BR04}, optimal $p_{\textrm s}=1/2$~\cite{GKE06} \\
    \hline %\br
  \end{tabular}
%  \end{indented}

  \caption{The four quantum 
  gates described in the text. 
  $(d_{\textrm s}, d_{\textrm f})$ denote 
  the number of edges gained on success and 
  the number of edges
  deleted per chain on failure, respectively. $p_{\textrm{NDM}}$ 
    is the probability of success 
    of a photon number non-demolition measurement that has to follow 
    the respective gate.\label{tab:gates} }
\end{table}

For other elementary probabilities, $p_{\textrm s}>0$, 
one may
ask:  What is the method of choice of dealing with
the intrinsic randomness? 
This is the question that we address in this work.
The analysis presented in this work
complements recent analytical
investigations~\cite{KGE06,GKE06} and 
numerical work on this topic~\cite{RB07} (see also 
Ref.\ \cite{Duan}).
In contrast to Refs.\ \cite{KGE06,GKE06}, where also rigorous asymptotic bounds have been presented, 
we here solely investigate the optimal and worst strategies 
for finite $N$ using a computer-assisted proof,
for arbitrary $p_{\textrm s}$ and different gates.
We discuss also the algorithm which is capable of finding the
provably optimal classical control strategy and delivers symbolic
results for the expected final size of the cluster.
Finally, some detailed numbers on the resource 
consumption in the preparation of 2D cluster states 
are given.

\section{Techniques}

In the first part of this work 
we aim at building up linear cluster chains from a 
reservoir of $N$ maximally entangled 
pairs. Our interest lies in identifying
the optimal classical control strategy
-- hence it will be fruitful to abstract from the underlying quantum
system. Indeed, at a given point in the process of building up the
cluster, we have a collection of cluster chains of various lengths at
our disposal. 
The ``state'' of our system
can hence solely be described by the respective lengths of the 
chains. This information will be represented by 
a \emph{configuration} vector $c= (c_1,c_2,\dots, c_{\text{max}})$, 
where $c_i$ is the number
of chains of length $i$, as counted by the number of 
\emph{edges}. An EPR pair, for example,  
has hence length $1$. The question of optimal
cluster state preparation
then becomes essentially a combinatorical one.
%
%Questions of imperfections come
%into play when one asks about storage times and errors in the 
%actual preparation \cite{Rohde}.

For the task of joining intermediate cluster states 
to form longer chains, we employ entangling gates 
like the Type-I fusion gate~\cite{BR04}, which 
plays a prominent role in linear optical architectures.
We restrict ourselves to two 
qubit gates that act symmetrically and with the same action for 
chains of all lengths -- most suggested quantum gates do have this property. The effective actions on two chains of 
lengths $l_1$ and $l_2$ in the number of edges
can then be described
by the outcomes on {\it success} and {\it failure}, 
\begin{eqnarray}	
	\{l_1,l_2\} &\mapsto  & \{l_1+l_2+d_{\textrm s}\},\\
	\{l_1,l_2\} & \mapsto  & \{l_1-d_{\textrm f},l_2-d_{\textrm f}\},
\end{eqnarray}
respectively, and are cast into the tuple $(d_{\textrm s}, d_{\textrm f})$.
This family embodies 
four gates: On the one hand it contains two
with an undefined failure outcome. To obtain a proper cluster state,
additional $Z$-measurement on the neighbors have 
to cut off the end qubits,
thus $d_{\textrm f}=2$. Additionally, there are those gates with 
a ``built-in'' $Z$-measurement on failure, $d_{\textrm f}=1$. 
On the other hand, there are two ``parity check'' gates (no 
new edges are created, $d_{\textrm s}=0$) 
and controlled-$Z$ (entanglement is created, $d_{\textrm s}=1$). See table~\ref{tab:gates} for more details.

\begin{figure}[t]
  \includegraphics{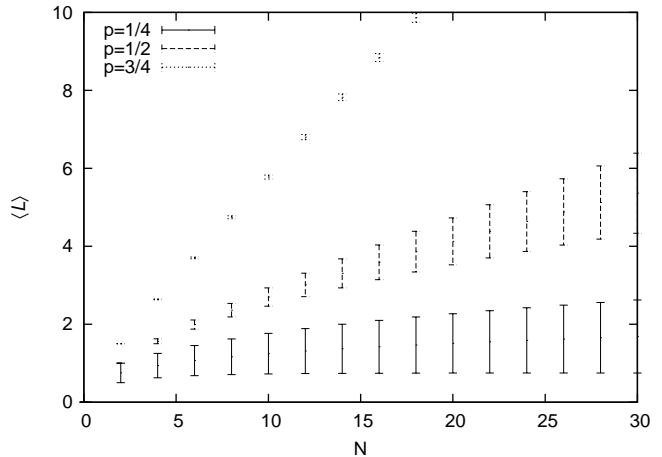}
	\caption{The 
	graphs show the influence of the classical strategy on the
	 expected length of the final cluster for Type-I fusion gates
	 operating at $p_{\textrm s}=1/4, 1/2$ and $3/4$, respectively.
	 For each probability, the range enclosed by the error bars
	 indicates the spread between the best possible and worst possible
	 strategy.
	\label{fig:fusion} }
\end{figure}

\begin{figure}[b]
  \includegraphics{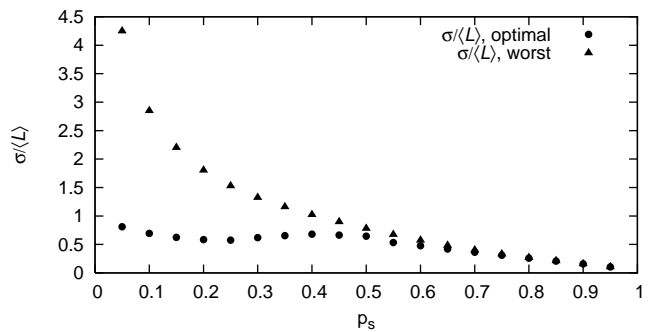}
	\caption{Relative standard deviation $\sigma/\langle L\rangle$ of
	the distribution of final chains for the optimal and worst strategy
	respectively. 
	It turns out that for high success probabilities, the statistical
	fluctuations for a single fixed strategy are more pronounced than
	the difference between even the best and the worst strategy.
	%For comparison the difference between the average final length of
	%the optimal and worst strategies relative to the average performance
	%of the two is shown.  The two bounds converge even faster than the
	%variances vanish, even the distributions themselves do not vary much
	%for high $p_{\textrm s}$.  
	\label{fig:deviations} }
\end{figure}

We can now describe our model for the generation of cluster states.
The 
starting point is a configuration consisting of $N$ EPR pairs. In each
step of the process, we choose two chains out of the repository and
try to fuse them.  The choice which specific two chains to
take is determined by the classical strategy.  A {\it
strategy} is
hence a complete  prescription which chains to fuse for all possible
configurations that may occur. This is thought to be a sequential
process. Note that if certain parts of a configuration are certainly
not used as resources in other steps, one can 
practically perform some of the steps in parallel. For the
theoretical analysis of resource consumption, however, it is 
always legitimate to think of a sequence of gates that is applied.

The process continues until either i)
there is only a single chain left or ii) the strategy decides to 
keep only the longest chain in the current configuration and 
to disregard the rest.
Note that for the case $p_{\textrm s}=1/2$ -- 
of central importance in linear optical architectures --
one never benefits from halting before all smaller chains have
been consumed~\cite{GKE06}. 
The performance of the strategy can 
then be measured by the expected length of the longest chain in the 
final configuration: Since this is a probabilistic process, the 
outcome will in general be an expectation value $\langle L\rangle$ 
of the final
length over all possible final configurations. 
Note that there are \emph{two different} means involved here: On the
one hand, there is the distribution of cluster states of different
lengths in a given configuration. On the other hand, since 
cluster state preparation is a
probabilistic process, there is the distributions of configurations
in the first place. It is the role of these two different 
means that renders
the discussion of the influence of the classical strategy involved.

We will obviously 
be interested in the performance of the optimal strategy,
\begin{equation}
	Q=\langle L\rangle_{\rm opt}(N), 
\end{equation}
which is the final average chain
length under the optimal strategy, 
which we will also refer to as the {\it quality}
of the configuration consisting 
of $N$ EPR pairs. We will also pay quite a significant
attention to the worst strategy, however, to assess
what influence the decision process can possibly
have. More concretely, we will ask what are the longest and shortest expected
lengths one can obtain out of $N$ initial EPR pairs, if one continues to
try to fuse all intermediate resources together until only a single
chain is left.

\section{Arbitrary gate probabilities}

Depending on the physical context, $p_{\textrm s}$ can vary over 
quite a range of different values.
In the {\it linear optical context}, 
$p_{\textrm s}=1/2$ plays a prominent role as the
optimal success probability of the parity check gate~\cite{GKE06}. 
When taking inefficient detectors  into account, needless to say,
this success probability will rapidly decrease below the 
theoretical value of $p_{\textrm s}=1/2$.

In Ref.\ \cite{Monroe}, probabilistic quantum gates on remote 
{\it trapped atom qubits} have been considered, exploiting 
interference of optical frequency qubits.
Here, the success probability 
$p_{\textrm s}$ is relatively small,  
smaller than $1/4$ even for perfect photon 
detectors.
However, the respective measurement 
gate
% -- effectively acting like a type-I fusion gate when considered
%in the appropriate local basis --
is constructed in a way to be 
very robust with respect to noise. We will see that for these kind of gates
operating at a relatively low probability of success, the choice of strategy
is in fact crucial.

\begin{figure}[b]
  \includegraphics{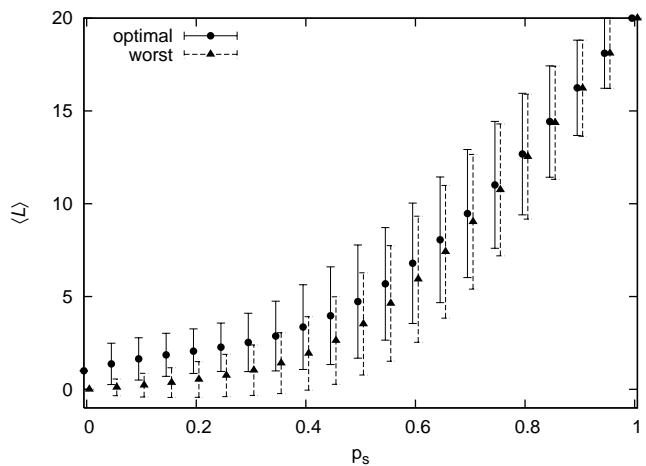}
  \caption{Expected final length of the optimal and the worst strategy, 
  starting from $N=20$ EPR pairs. 
  Error bars give the standard deviation of the final length 
  distributions. To allow comparison but ensure distinguishability at the same time
  the data have been shifted by $0.005$ to the left and to the right,
  respectively.
  \label{fig:errorbars} }
\end{figure}

However, gates constructed using {\it weak non-linearities}
in optical systems as considered in 
Refs.\ \cite{LNMS06}
can result 
in $p_{\textrm s}=1-2^{-1-n}$
where $n$ is the number of ancillas used~\cite{LNMS06}. In this context, we will see
that the worst strategy performs essentially 
identical to the optimal strategy, hence
confirming that in the regime of high success probability, strategic choice 
hardly matters. 

In contrast to other approaches, 
we give exact values for the optimal and 
the worst strategy. Furthermore, we will stick
to our previous notion of strategies acting on a fixed reservoir of resource states. However, using
Theorem~16 from Ref.\ 
\cite{GKE06}, these results also deliver the solution to the 
converse question, of the average resource consumption for construction of a chain of given length
(see Ref.\ \cite{RB07} for numerical results on this 
converse question, namely the generation of fixed length
chains using an infinite supply of entangled pairs).

\begin{figure}[t]
  \includegraphics{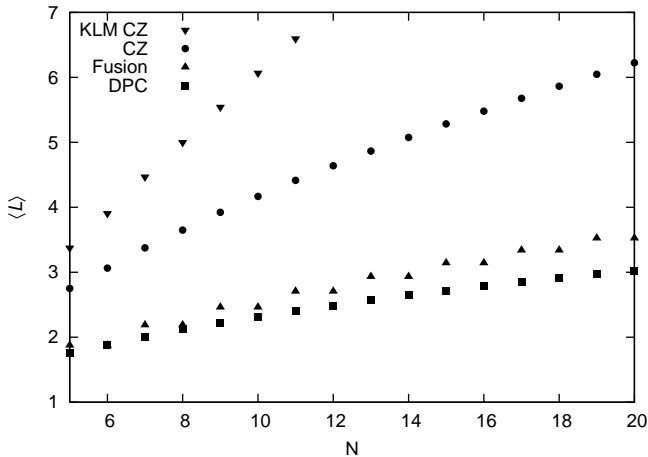}
	\caption{The optimal expected length $\langle L \rangle$
	of the final cluster for the gates listed in Table
	\ref{tab:gates}. For comparison all gates are assessed at
	$p_{\textrm s}=1/2$.
	\label{fig:quality} }
\end{figure}

Figure~\ref{fig:fusion} shows how the optimal and the worst strategy 
perform when using the Type-I fusion gate at different probabilities. 
Examples from three different {\it regimes} 
are shown, the distinctness of which is quite 
remarkable:

\begin{itemize}
  \item For small gate probabilities the choice of 
  strategy is crucial.
  Applying gates in a particularly unfortunate 
  fashion might actually
  result in no net increase of the
    length of the longest chain at all. However, employing the optimal
    strategy 
    the length of the longest chain turns out to be 
    an increasing function of the number of EPR pairs,
    rather than a constant one. For the asymptotic behavior, this is 
    actually shown in Ref.\ \cite{GKE06}.
    
  \item In some intermediate regime 
    (like $p_{\textrm s}=1/2$ for Type-I fusion gate) both
    the optimal as well as the 
    worst strategy result in a length increase of the longest chain.
    But still, the choice of strategy distinguishes whether an efficient growth, i.e.,
    a linear growth $O(N)$ will be obtained, in constrast
    to merely a growth of $O(\sqrt{N})$ (shown in Ref.~\cite{GKE06}). 
         
  \item What may well be considered intuitively plausible is confirmed 
    by the simulations as well: For $p_{\textrm s}$ being large, 
    the difference in the performance of strategies
    becomes negligible for practical purposes. The order in which the fusion attempts are
    carried out hence hardly matters. This comes in handy as one does not 
    have to concentrate on the optimal strategy
    in experimental realizations any
    more. Even though the optimal strategy 
    also realizes the least storage time in this setting, 
    there is also almost no difference in the
    storage time required by the different strategies. 
    Then, of course other meaningful figures of merit enter center
    stage, like the amount of feed-forward or rerouting needed.
    Strategies like {\sc Static}~\cite{GKE06} that only fuse nearest neighbors in the repository of chains will then be favorable as 
    they realize the least amount of feed-forward in
    this setting simultaneously. This strategy is already heavily parallelized,
    in the sense that many fusion operations are applied in parallel rather
    than waiting after each fusion for the outcome of the previous attempt,
    decreasing the overall preparation time.
    Of course, this is another property which is highly desirable and attention can now
    be centred upon it.

    Minimizing the average 
    storage time as well as the number of applications of the entangling gate
    are crucial to prevent the system from unnecessary strong decoherence. Depending on the physical implementation
    a suitable figure of merit in this context could also be the 
    storage time weighted with the chain size.
\end{itemize}
In this latter setting, there is no need for shuffling chains around,
rather than only single qubits. 
A huge amount of rerouting on the
level of qubits is still needed. Ref.~\cite{KRE06} 
presents a possible solution to this problem, making use of
tools from percolation theory.
    
Figures~\ref{fig:deviations} and~\ref{fig:errorbars} show the dependence of 
the standard deviation $\sigma$ of the final length distributions on the gate probability.
As the gate probability increases not only the upper and lower bound to strategy 
performances converge, but also their relative variances converge and vanish. 
Interestingly enough,
the relative difference between the two strategies 
gets even smaller than their relative standard deviations.

\begin{figure*}[t]
  \includegraphics{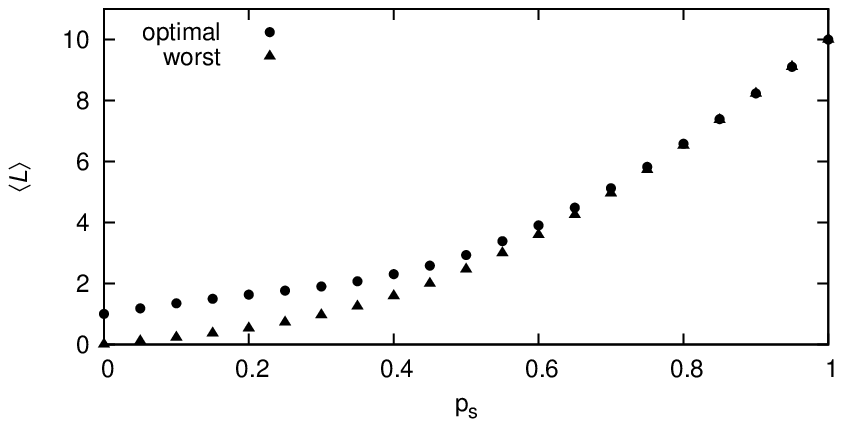}%
  \includegraphics{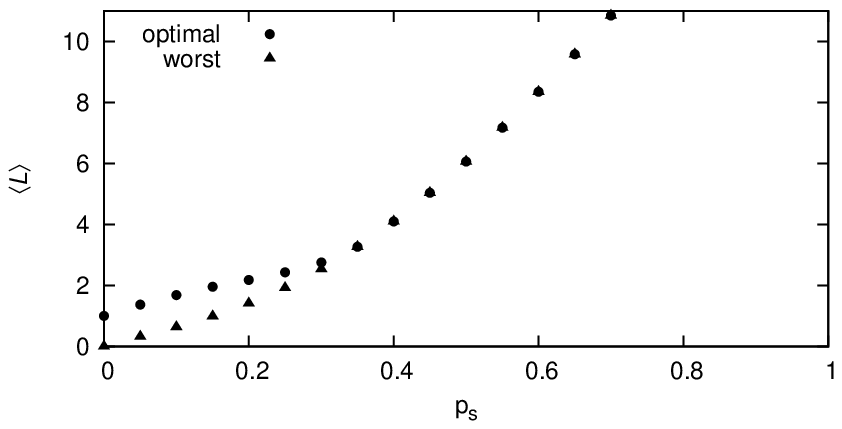}
	\caption{Performance of the optimal and 
	worst strategy for $N=10$
	for Type-I fusion (left graph) and the 
	KLM controlled-Z gate (right
	graph). Both gates have the
	same number of lost edges on failure. \label{fig:gatesp} }
\end{figure*}

\section{Additional gates}

The action of all gates used here can be described by a pair of 
two parameters $(d_{\textrm s},d_{\textrm f})$ (similar to the treatment in Ref.\ \cite{RB07}): 
The number of edges $d_{\textrm f}$ that are deleted 
from the participating chains in case of failure and
the number of edges  $d_{\textrm s}$
that are added to the resulting chain when the gate is successful.
The Type-I fusion gate deletes the participating qubits in case of failure. 
The length of the resulting chain on success is the sum of the lengths of the 
original chains. Therefore, this gate can be described by the parameters $(1,0)$.
Besides having investigated its performance at different success probabilities,
we further compared it to other possible entangling gates.
A controlled-$Z$ gate (CZ) 
creates an edge between the qubits it is acting on. 
However, if on failure the outcome is 
not known~\cite{RB07}, the two
qubits have to be deleted 
by applying $Z$-measurements to the neighboring sites. 
Thus, this gate then 
effectively deletes $2$ edges from both chains on
failure, and it is represented by $(2,1)$. The linear optical implementation of the CZ gate (KLM CZ gate~\cite{KLM01})
has, in contrast, 
a defined error outcome which is effectively 
the action of a $Z$-measurement on the two
qubits. Therefore, this gate simply cuts the two qubits from 
their chains on failure, and consequently
is characterized by $(1,1)$.
For the sake of completeness we also introduce a gate which 
does not add any new edges (like the fusion gate), but deletes two 
edges from each chain on
failure (like the CZ gate), thus resulting in 
$(2,0)$. This gate might be a parity check gate, like the 
fusion gate, but with an undefined failure outcome,
thus denoted by DPC (destructive parity check). Actually,
the individually trapped atoms can be used to implement this type
of gate~\cite{Monroe}.

Further gates would be for example the
Type-II fusion~\cite{BR04}, or the one created
by weak non-linearities using the qubus technique~\cite{LNMS06}. 
However, the first one is hard to
compare with the gates presented here as it requires three qubit cluster states
(locally equivalent to GHZ states, $|0,0,0\rangle+|1,1,1\rangle$),
rather than EPR pairs to start with.
The latter one is actually already included in our analysis, as 
it corresponds to a gate with parameters $(1,1)$, operating with a
gate-probability $p_{\textrm s}=3/4$.

The performance $\langle L\rangle_{\rm opt}$ of the 
optimal strategy using all four 
gates is shown in figure~\ref{fig:quality}.
The difference between the gates in the number of edges consumed on 
success and on failure is clearly reflected
by the optimal average final length that is achieved. As noted earlier, we do observe
an interesting crucial dependence of the performance of the 
best and worst strategy while varying $p_{\textrm s}$:
$1/2$ marks some {\it intermediate regime} for the Type-I fusion gate, 
above which the choice of strategy does not
carry weight anymore. This behavior is indicated by the 
diminishing gap in the first part of figure~\ref{fig:gatesp}. The 
second part
shows this plot using the KLM CZ gate. Interestingly, in 
this case the worst and optimal strategy are already 
indistinguishable at gate probabilities
far below $1/2$. Not only the probability, but also 
which type of gate is used influences how important the 
classical choice of strategies will be.

\section{Algorithm}

\begin{figure*}[t]
\hspace{-\parindent}
\hbox to \hsize{\begin{minipage}[t]{\columnwidth}
\rule{\columnwidth}{0.5pt}
\raggedright
\begin{tabular}{ll}
{\sc Name:}		& $\operatorname{Optimize}$ \\
{\sc Input:} 	& Integer $n$ \\
{\sc Output:}	& For all configurations $C$ with up to $n$ vertices,\\ 
              & the global variable $Q(C)$ is set to the quaity of $C$.
\end{tabular}
\rule{\columnwidth}{0.5pt}
\begin{tabbing}
	\quad\quad\=\quad\=\quad\=\quad\=\quad\=\quad\=\quad\=\quad\= \\
	1\> $\operatorname{SUB\ Optimize}(n)$ \\
	2\>\> $\operatorname{for} i:=1 \operatorname{to} n$ \\
	3\>\>\> $C:=\operatorname{AllConfs}(n)$ \\
	4\>\>\> $\operatorname{foreach} c \in C$ \\
	5\>\>\>\> $\operatorname{OptimizeConf}(c)$ \\
	6\>\>\> $\operatorname{end\ foreach}$ \\
	7\> $\operatorname{end\ for}$ 
\end{tabbing}
\rule{\columnwidth}{0.5pt}
\end{minipage}

\begin{minipage}[t]{\columnwidth}
\rule{\columnwidth}{0.5pt}
\raggedright
\begin{tabular}{ll}
{\sc Name:}		& $\operatorname{OptimizeConf}$ \\
{\sc Input:} 	& Configuration $c$\\
{\sc Assumption:} & For all configurations $c'$ with fewer particles than $c$, \\
							 & the global variable $Q(c')$ is set to the quality of
							 $c'$. \\
{\sc Output:}	& Sets global variable $Q(c)$ to quality of $c$.
\end{tabular}
\rule{\columnwidth}{0.5pt}
\begin{tabbing}
	\quad\quad\=\quad\=\quad\=\quad\=\quad\=\quad\=\quad\=\quad\= \\
	1\> $\operatorname{SUB\ OptimizeConf}(c)$ \\
	2\>\> $l:= \operatorname{length(c)}$ \\
	3\>\> $\operatorname{for} i\leq l$ \\
	4\>\>\> $\operatorname{for} j<i$ \\
	5\>\>\>\> $q[i,j]:=p_s\, Q(\operatorname{fuse}(c,i,j)) +
	(1-p_s) Q(\operatorname{fail}(c,i,j))$ \\
	6\>\>\> $\operatorname{end\ for}$ \\
	7\>\> $\operatorname{end\ for}$ \\
	8\>\> $Q(c):=\operatorname{max}_{i,j} q[i,j]$ 
\end{tabbing}
\rule{\columnwidth}{0.5pt}
\end{minipage}
}

\caption{The recursive algorithm which computes the optimal expected
length one can obtain from a given configuration. It relies on three
simple sub-routines: 
$\operatorname{AllConfs}(n)$, which returns a
table of all possible configurations with up to $n$ particles; 
$\operatorname{fuse}(c,i,j)$, returning the configuration resulting
from $c$ after a successful fusion of the $i$-th and $j$-th chain and
finally $\operatorname{fail}(c,i,j)$ which acts likewise, but assumes
the fusion to fail.
\label{fig:algorithm}}
\end{figure*}

Can one use a computer-assisted proof to identify the optimal
strategy for a given configuration? Naively, one would expect this not
to be the case, as the number of configurations with $n$ entangled
bonds is exponential in $n$~\cite{GKE06} and the number of strategies is, in turn,
exponential in the number of configurations. A brute-force search
for the optimal strategy is clearly unfeasable. Fortunately, the
problem can be addressed using a smarter, recursive algorithm, 
as a variant of a backtracking algorithm.

To understand how, note that as a result of an attempted fusion, the
number of entangled particles decreases by one in case of success and
by at least two in case of failure. Now consider a given configuration
$c$ with $n$ vertices (not edges) 
and assume that we know the quality $Q$ for all
configurations of strictly fewer than $n$ constituents. If the number
of chains in $c$ equals $l$, then there are roughly $l^2$ possible
choices a strategy can make. If the strategy decides on fusing chains
$i$ and $j$, then the expected final length is going to be
\begin{equation}
  p_s\, Q(\operatorname{fuse}(c,i,j)) + (1-p_s)
	Q(\operatorname{fail}(c,i,j)).
\end{equation}
The preceeding value can be computed explicitly, 
because by assumption $Q$ is known 
for both $\operatorname{fuse}(c,i,j)$ and
$\operatorname{fail}(c,i,j)$. Hence, using $O(l^2)$ tests, one can
identify the optimal pair $(i,j)$.

A computer implementation builds up a lookup table, i.e., 
a list containing the quality $Q$ of every configuration up to a given number
of particles $n$. The process starts with the trivial
configuration $n=2$, for which the quality is known and works its way
up to higher $n$ as described before. Note that the optimal
strategy can only be specified by means of a lookup table for
all configurations, and there is generally no ``easy
description'' of the optimal strategy, 
so a description with a small Kolmogorov complexity. 

Figure~\ref{fig:algorithm} presents 
an explicit pseudo-code implementation. By adjusting the parameter 
$p_s$ and the subroutines
$\operatorname{fail}()$ and $\operatorname{fuse}()$, the program is easily
modified to general gates. Also note that finding the 
\emph{worst} possible strategy can be achieved similarily.

The usefulness of practical implementations will be limited by memory
consumption, as the lookup table grows linearly in the number of
configurations, which is exponential in $n$. We utilized the computer
algebra system \emph{Mathematica} to assess the quality of
configurations consisting of several dozen particles. A desktop
computer completes the calculations in a few hours.

\begin{figure}[t]
  \includegraphics{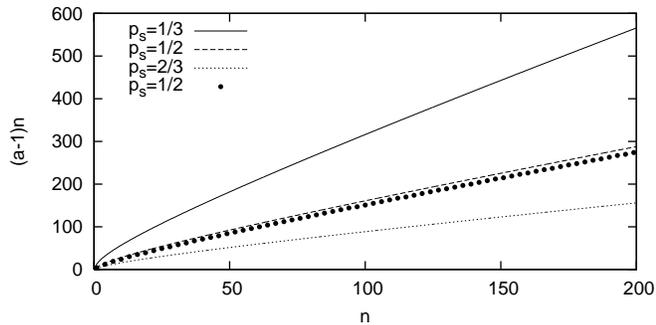}
  \caption{Overhead $(a-1)n$ in the weaving provess with elementary probability $p_{\textrm s}=1/3,1/2,2/3$
    to succeed for cluster size $n$ with a probability of at least $P_{\textrm s}=19/20$.
    In addition to the upper bounds, exact values for $p_{\textrm s}=1/2$ are given. \label{fig:2d_an} }
\end{figure}

The actual optimal strategy does in fact not have to be unique. However, our
investigations indicate that taking the smallest chains most of the time is optimal.
Deviations from this rule depend on the success probability and there is no obvious way
to describe them with such a small complexity as ``always take the shortest chains''.
For some insight in these deviations see Ref.~\cite{GKE06}.
However, a worst strategy seems to be, regardless of the success probability, the strategy
that always acts on the longest pair of chains.

\section{Preparation of 2D structures: ``weaving''}

We now turn to the preparation of two-dimensional structures, universal for
quantum computing. 2D cluster states can be generated starting from linear chains.
This can be achieved by placing them on top of 
each other so that they inherit the square lattice geometry
of the cluster states, and by applying entangling gates at the crossings subsequently. 
This procedure ({\it weaving}) obviously gives rise to the required 2D structure when
all gates succeed. However, sufficiently many 
spare qubits have to be available for each crossing in case a gate fails.

Here, the probabilistic character of the employed quantum gates is again crucial:
One has to make sure that the preparation becomes \emph{almost certain} 
for large 2D cluster states. Fortunately, it turns out that this aim can always be achieved: 
As has been shown in Ref.\ \cite{GKE06} 
(compare also Refs.\ \cite{Duan} for a strong indication
suggesting a polynomial bound), starting from linear cluster chains, 
2D cluster states of arbitrary size $n\times n$
can be built almost certainly for large $n$,
with an overhead per site that only depends on the gate probability.
Hence, the overall probability of success satisfies
\begin{equation}
	P_{\textrm s}(n)\rightarrow 1, 
\end{equation}
when
using $O(n^2)$ EPR pairs for a state of size $n\times n$. Moreover, this is true 
for any $p_{\textrm s}>0$.  This is obviously
the optimal scaling that can be achieved for 2D cluster states \cite{GKE06}.

The gates necessarily act on vertices 
located along the chain, and not only on vertices at the end of the chain. 
Hence, the employed quantum gates must provide suitable
error outcomes in order not to tear the chains apart. 
Then, the weaving procedure as proposed in Ref.\ \cite{KGE06} can 
be used to construct a cluster using gates succeeding with any 
$p_{\textrm s}>0$ almost certainly with a constant overhead per site.
This procedure consists of $n$ parallel chains in the horizontal direction
and one long chain that connects the short ones in the vertical direction.
To find the actual prefactor of the leading quadratic term, we recall 
the processes that consume edges when
constructing a site:
\begin{itemize}
  \item The cluster construction itself -- there are $n^2$ sites and $2n(n-1)$ edges.
  \item A $\sigma_z$ measurement is applied to a qubit on one of the chains, destroying $2$ edges.
  \item Failures of the fusion gate result in alternating deletions of two edges from the two chains involved.
\end{itemize}
Fixing an overall success probability results in a number of overhead edges that determines the number of
possible failures. Up to a constant error per chain -- thus only linear in $n$ -- 
the number of failures equals the number of edges consumed. The number of 
edges per chain depends linear on
the side length of the cluster to be produced, the coefficient being 
$a$ (see figure~\ref{fig:2d_an}).

\begin{figure}[t]
  \includegraphics{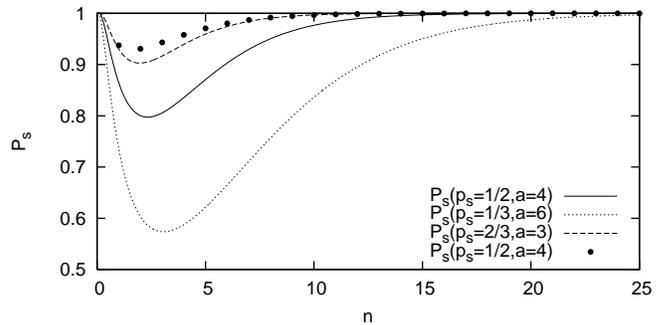}
  \caption{The weaving success probability $P_{\textrm s}( n )$ for fixed gate probability $p_{\textrm s}=1/3,1/2,2/3$ and overhead $a=2/p_{\textrm s}$.
    For $p_{\textrm s}=1/2$ exact values are given for comparison with the lower bound. \label{fig:2d_pn} }
\end{figure}

\begin{figure}[b]
  \includegraphics{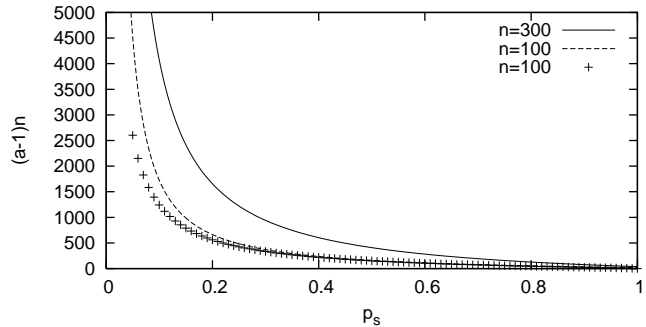}
  \caption{Overhead $(a-1)n$ that is needed 
  for the weaving process to succeed for a fixed cluster size $n=100,300$ with a probability of at least 
  $P_{\textrm s}(n)=19/20$.
    For comparison with the upper bounds, exact values are given for $n=100$ as well. \label{fig:2d_ap} }
\end{figure}

The lower bound for the overall success probability~\cite{GKE06}
\begin{equation}
  P_{\textrm s} (n) 
  \ge \left( 1 - \exp\left( -\frac{2(anp_{\textrm s}-n+1)^2}{an} \right) \right)^n
  \label{eqn:P}
\end{equation}
is displayed in figure~\ref{fig:2d_pn}. Chosing $a>1/p_{\textrm s}$ will result in $P_{\textrm s}(n) \rightarrow1$ as
$n\rightarrow\infty$. By fixing $n$ and $p_{\textrm s}$ we can extract from~(\ref{eqn:P}) $a$
for a given $P_{\textrm s}$, 
so also the number of overhead edges
that are required to achieve at least
this overall success probability,
\begin{equation}
  a = 1/p_{\textrm s} + \varepsilon + o(1)
\end{equation}
with $\varepsilon > 0$. Figure~\ref{fig:2d_ap} shows the dependence of the 
required resources on the gate probability $p_{\textrm s}$.
Summing up all these contributions we arrive at a resource consumption when
building 2D clusters from existing chains of at least
\begin{equation}
  4+2(1/p_{\textrm s}-1)
\end{equation}
per site, for example $9$ edges at $p_{\textrm s}=1/2$.

This prefactor -- so far the best known one -- can possibly be improved still. 
For example, one may aim at not preparing a 2D
cluster state, but a state that is equivalent to such a state, up to local unitary rotations.
Specifically, one could aim at preparing a graph state that is equivalent to a cluster up to local Clifford
operations. The orbit under such local Clifford operations is reflected on the level
of graphs by the orbit under \emph{local complementations} 
\cite{Graph1,Graph2}. Ref.\ \cite{MIT} already exploit
such local complementations when efficiently 
preparing 2D structures. It would be interesting to see
whether a systematic explorations of these tools give rise to a 
significant improvement of the above prefactor in the optimal quadratic scaling.

\section{Summary}

We have applied the tools introduced in Refs.\ 
\cite{KGE06,GKE06} to a number of different quantum gates to prepare cluster 
states for quantum computing from
elementary blocks. Depending on the underlying physical architecture, the gates
operate at different success probabilities.
A qualitatively different behavior of the difference between the optimal and the worst outcomes 
has been  observed when varying $p_{\textrm s}$. The specific probability at which 
this transition occurs depends on the parameters of the gate used,
i.e., how many edges are consumed on success and failure. 
At low probabilities the choice of strategy is highly significant, 
and any scheme for building up cluster states 
has to be based on a good choice of a strategy.
For gates at high $p_{\textrm s}$, the potential difference is negligible. Hence, in this regime, 
other figures of merit, minimizing imperfections and error proparation~\cite{Rohde} in an actual experimental
context, become the relevant 
key quantities. This work provides a guideline to what 
extent the choice of the
classical preparation strategy is crucial. Similar ideas could also be applied when building up structures that 
may be used for fault-tolerant or error resilient schemes \cite{FT,FT2}. 
Needless to say, we have concentrated on the preparation of
cluster states for measurement-based quantum computing. It would 
be interesting to see how the
new freedom of quantum computing using universal resources different from cluster states
as in Ref.\ \cite{New} affects the strategies of preparation.
 
\section*{Acknowledgements}

This work has been supported by the DFG, the EU (QAP), the EPSRC, QIP IRC, Microsoft
Research through the European PhD Scholarship Programme, and the EURYI award scheme.

%\section*{References}


\begin{thebibliography}{99}

\bibitem{BR01}
    R.\ Raussendorf  and H.J.\ Briegel,
    {Phys.\ Rev.\ Lett.} {\bf 86}, 910 (2001);
    H.J.\ Briegel and R.\ Raussendorf,
    {\it ibid.} {\bf 86}, 5188 (2001).

\bibitem{Survey}
       D.E.\ Browne and H.-J.\ Briegel, quant-ph/0603226;
               M.\ Hein,
        W.\ D{\"u}r, J.\ Eisert,
        R.\ Raussendorf, M.\ Van den Nest, and
        H.-J.\ Briegel,  quant-ph/0602096.

\bibitem{New}
	D.\ Gross and J.\ Eisert, quant-ph/0609149.
	
\bibitem{Cold}
 	D.\ Jaksch, H.-J.\ Briegel, J.I.\ Cirac, 
	C.W.\ Gardiner, and P.\ Zoller,
	Phys.\ Rev.\ Lett.\ {\bf 82}, 1975 (1999).
		
\bibitem{Greiner}
	O.\ Mandel, M.\ Greiner, A.\ Widera, T.\ Rom, T.W.\ H{\"a}nsch, 
	and I.\ Bloch, Nature {\bf 425}, 937 (2003).

\bibitem{KLM01}
    E.\ Knill, R.\ Laflamme, and G.J.\ Milburn,
    {Nature} {\bf 409}, 46 (2001).
    
\bibitem{BR04}
    D.E.\ Browne and T.\ Rudolph,
    {Phys.\ Rev.\ Lett.} {\bf 95}, 010501 (2005).

\bibitem{Zeilinger}
	 P.\ Walther, K.J.\ Resch, T.\ Rudolph, E.\ Schenck, 
	 H.\ Weinfurter, V.\ Vedral, M.\ Aspelmeyer, and A.\ 
	 Zeilinger, Nature {\bf 434}, 169 (2005).

\bibitem{BCZ+06}
	X.-H.\ Bao, T.-Y.\ Chen, Q.\  Zhang, J.\ Yang, H.\ Zhang, 
	T.\ Yang, and J.-W.\ Pan, quant-ph/0610182.

\bibitem{GKE06}
    D.\ Gross, K.\ Kieling, and J.\ Eisert,
    {Phys.\ Rev.\ A} {\bf 74}, 042343 (2006).

\bibitem{LNMS06}
    S.G.R.\ Louis, K.\ Nemoto, W.J.\ Munro, and T.P.\ Spiller,
    New J.\ Phys.\ {\bf 9}, 193 (2007); 
    Phys.\ Rev.\ A {\bf 75}, 042323 (2007).

%\bibitem{Nonlinear2}
%	T.P.\ Spiller, K.\ Nemoto, S.L.\ Braunstein, 
%	W.J.\ Munro, P.\ van Loock, and 
%	G.J.\ Milburn,
%	New J.\ Phys.\ {\bf 8}, 30 (2006).
	
%\bibitem{Nonlinear3}
%	J.H.\ Shapiro and M.\ Razavi, quant-ph/0612086.	

\bibitem{Atoms}
	C.\ Cabrillo, J.I.\ Cirac, P.\ Garchia-Ferndandez, and 
	P.\ Zoller, Phys.\ Rev.\ A {\bf 59}, 1025 (1999).
		
\bibitem{Monroe}
	L.-M.\ Duan, M.J.\ Madsen, D.L.\ Moehring, 
	P.\ Maunz, R.N.\ Kohn Jr., and 
	C.\ Monroe, quant-ph/0603285. 

%\bibitem{Cavity2}
%	I.E.\ Protsenko, G.\ Reymond, N.\ Schlosser, and P.\ Grangier, 
%	Phys.\ Rev.\ A {\bf 66}, 062306 (2002).

\bibitem{Kimble}	
	L.M.\ Duan and H.J.\ Kimble, 
	Phys.\ Rev.\ Lett.\ {\bf 90}, 253601 (2003).
	
\bibitem{Cavity3}	
	D.E.\ Browne, M.B.\ Plenio, and S.\ Huelga, 
	Phys.\ Rev.\ Lett.\ {\bf 91}, 067901 (2003).
	
\bibitem{Cavity4}
	Y.L.\ Lim, S.D.\ Barrett, A.\ Beige, P.\ Kok, and 
	L.C.\ Kwek,
	Phys.\ Rev.\ A {\bf 73}, 012304 (2006).

\bibitem{Scheel}
	S.\ Scheel and N.\ L{\"u}tkenhaus, 
  	New J.\ Phys.\ \textbf{6}, 51 (2004).

\bibitem{Eisert05}
    J.\ Eisert,
    {Phys.\ Rev.\ Lett.} {\bf  95}, 040502 (2005).
			
\bibitem{KGE06}
    K.\ Kieling, D.\ Gross, and J.\ Eisert,
    {J.\ Opt.\ Soc.\ Am.\ B} {\bf 24}, 184 (2007).
		
\bibitem{RB07}
    P.P.\ Rohde and S.D.\ Barrett,
    New J.\ Phys.\ {\bf 9}, 198 (2007).

\bibitem{Duan}
	L.-M.\ Duan and R.\ Raussendorf, 
	Phys.\ Rev.\ Lett.\    {\bf  95}, 080503 (2005).

\bibitem{PJF01}
    T.B.\ Pittman, B.C.\ Jacobs, and J.D.\ Franson,
    {Phys.\ Rev.\ A} {\bf 64}, 062311 (2001).
	
\bibitem{RLBW02}
    T.C\ Ralph, N.K.\  Langford, T.B.\  Bell, and A.G.\ White,
    {Phys.\ Rev.\ A} {\bf 65}, 062324 (2002).

\bibitem{HT02}
    H.F.\ Hofmann and S.\ Takeuchi,
    {Phys.\ Rev.\ A} {\bf 66}, 024308 (2002).

\bibitem{Rohde}	
	P.P.\ Rohde, T.C.\ Ralph, and W.J.\ Munro,
	quant-ph/0701090.

\bibitem{KRE06}
    K.\ Kieling, T.\ Rudolph, and J.\ Eisert,
    quant-ph/0611140.

\bibitem{Chen}
	Q. Chen, J.\ Cheng, K.-L.\ Wang, and J.\ Du,
	Phys.\ Rev.\ A {\bf 73}, 012303 (2006).
	
\bibitem{Graph1}
	M.\ Hein, J.\ Eisert, and H.J.\ Briegel,
	Phys.\ Rev.\ A {\bf 69}, 062311 (2004).
		
\bibitem{Graph2}
	M.\ Van den Nest, J.\ Dehaene, and B.\ De Moor,
	Phys.\ Rev.\ A {\bf 69}, 022316 (2004).		
	
\bibitem{MIT}
	G.\ Gilbert, M.\ Hamrick, and Y.S.\  Weinstein, 
	Phys.\ Rev.\ A {\bf 73}, 064303 (2006).
 
\bibitem{FT}
	C.M.\ Dawson, H.L.\ Haselgrove, and M.A.\ Nielsen,
	Phys.\ Rev.\ Lett.\ {\bf 96}, 020501 (2006).
	
\bibitem{FT2}
	M.\ Varnava, D.E.\ Browne, and T.\ Rudolph, 
	Phys.\ Rev.\ Lett.\ {\bf 97}, 120501 (2006).
	
\end{thebibliography}
\end{document}